\documentstyle[12pt,aaspp]{article}
\begin{document}

\title{PROTON ACCELERATION BEYOND 100~EeV BY
AN OBLIQUE SHOCK WAVE IN THE JET OF 3C~273}

\author{Yasuko S. Honda\altaffilmark{1} and Mitsuru Honda \altaffilmark{2}}

\altaffiltext{1}{Department of Electrical and Information Engineering, Kinki 
University Technical College, Mie 519-4395, Japan; yasuko@ktc.ac.jp.} 
\altaffiltext{2}{Plasma Astrophysics Laboratory, Institute for Global Science, 
Mie 519-5203, Japan.} 

\begin{abstract}
We estimate the highest energy of proton diffusively accelerated by shock in
knot~A1 of the jet in luminous nearby quasar 3C~273.
Referring to the recent polarization measurements using very long baseline
interferometry (VLBI), we consider the shock propagation across magnetic
field lines, namely, configuration of the oblique shock.
For larger inclination of the field lines, the effects of particle reflection
at the shock front are more pronounced, to significantly increase
acceleration efficiency.
The quasiperpendicular shock turns out to be needed for safely achieving
the proton acceleration to the energy above $100~{\rm EeV}$
($10^{20}~{\rm eV}$) in a parameter domain reflecting conceivable
energy restrictions.
\end{abstract}

\keywords{acceleration of particles --- galaxies: jets --- magnetic fields --- methods: numerical --- quasars: individual (3C 273) --- shock waves}

\section{INTRODUCTION}

Core-dominated quasar 3C~273 is one of the most confirmed objects on account of
high optical luminosity and low redshift ($z=0.158$; Bridle \& Perley 1984).
The image obtained with the refurbished {\it Hubble Space Telescope} (HST)
revealed that the narrow optical jet consists of discrete knots
(Bahcall et al. 1995, hereafter B95) associated with shocks.
In radio bands the high linear polarization and featureless spectra
imply that the observed emission from the knots is originated in the
synchrotron radiation of electrons (and possibly positrons).
If the optical emission is also of the synchrotron origin,
{\it in situ} acceleration of electrons must be just taking place,
as they are compatible with the synchrotron lifetime inferred from
magnetic field strength (Meisenheimer et al. 1989: M89).
Polarization data in optical bands are similar to those in radio ones, and
suggest that the optical emission is attributed to the synchrotron losses of
accelerated electrons (R\"oser \& Meisenheimer 1991: RM91;
Conway et al. 1993: C93; R\"oser et al. 1996).
In the brightest knot~A, the observed power-law spectrum
($S_{\nu}\propto\nu^{-\alpha}$) can be fitted by the index of
$\alpha=0.45\pm 0.10$ (M89), close to the canonical value of
$\alpha=(\alpha_{\rm e}-1)/2\sim 0.5$ for the electron spectral index
$\alpha_{\rm e}\sim 2$, expected for Fermi acceleration
at strong nonrelativistic shocks.

As far as such an acceleration mechanism works for electrons, the same
mechanism will operate for acceleration of ions likewise, providing their
abundance in the jet is finite (e.g., Rawlings \& Saunders 1991).
One of the most promising mechanisms is the diffusive shock acceleration
(DSA) involving resonant scattering of injected particles, in which the
particles trapped in a mean magnetic field are resonantly scattered by the
Alfv\'en waves superposed on the mean field
(for review, Drury 1983; Blandford \& Eichler 1987;
Jones \& Ellison 1991; Longair 1992: L92), so as to migrate
back and forth between the upstream and downstream region of the shock.
As a consequence, the particles are accelerated,
receiving energy from the shock.
In this context, Biermann \& Strittmatter (1987) have discussed the
possibility of ultrahigh energy acceleration of protons.
However, their description was limited to a simple case for
parallel shock propagating along magnetic field lines.
Rachen \& Biermann (1993) estimated the highest proton energy
as $\sim 10^{21}~{\rm eV}$, but it seems to be rather optimistic. 

For precise evaluation of the highest energy, it is important to consider
the actual magnetic field geometry in the acceleration site.
With regard to this point, a helical pattern of the magnetized jet has
recently been observed by the Faraday rotation measures in the
{\it Highly Advanced Laboratory for Communications and Astronomy} VLBI
Space Observatory Programme (Asada et al. 2002), and it appears to
self-organize {\it double helical} structure (Lobanov \& Zensus 2001: LZ01).
In such a filamentary jet, expected is that huge current
($\sim 10^{16}~{\rm A}$: C93) generates
substantial magnetic fields transverse to the filaments
(Honda, Meyer-ter-Vehn \& Pukhov 2000; Honda \& Honda 2002: HH02).
Anyhow, when allowing shock propagation along the jet concomitant with
{\it non}-parallel components of the magnetic fields,
the shock likely crosses the field lines.
This corresponds to the configuration of the ``{\it oblique shock}'',
for which acceleration efficiency is higher than that for the parallel shock,
owing to the effects of mirror-reflection of particles at the shock front
(Jokipii 1987; Ostrowski 1988: O88).

In this Letter, we report a result of the numerical analysis for the highest
energy of proton diffusively accelerated by the oblique shock in knot~A1 of
3C~273 jet.
The present model of the oblique DSA is on the basis of the first-order
Fermi mechanism including the resonant scattering of particles
by the Kolmogorov turbulence.
Concerning quench of acceleration and energy loss processes,
more solid argument is expanded.
We demonstrate that for an expected range of radiation energy density,
the maximum proton energy exceeding $10^{20}~{\rm eV}$ is {\it safely}
achieved, especially for quasiperpendicular shock acceleration.

\section{THE MODEL OF DIFFUSIVE SHOCK ACCELERATION AT 3C~273/A1}
\subsection{\it Physical Properties of Knot~A1}

Below we outline the physical properties and parameters at the knot in the
leading edge of the jet of 3C~273, which is specified as ``knot A1'' (B95),
or ``knot A'' in earlier literatures.
The knot~A1 is well-resolved in optical wavelengths ($0.1''$ for HST),
and its radius is about $1~{\rm kpc}$ (R\"oser et al. 2000: R00).
This optical core is encased in the larger radio cocoon
whose half-width is $\sim 2~{\rm kpc}$ (B95); and
the overall structure seems to correspond to the ``hot spot''
(in the nomenclature for Fanaroff-Riley II sources; Rachen \& Biermann 1993),
which may be related to the upper limit of the size of shocks.
The key feature of this knot is that X-ray flux is more prominent
than that observed in the other knots (Marshall et al. 2001: M01).
Extrapolating the radio-to-optical spectrum to high-energy region
approximately reproduces the observed X-ray continuum (R00).
Accordingly, for the X-ray as well, the power-law spectrum
(with index $\alpha=0.60\pm 0.05$; M01) can be explained by the
electron-synchrotron model (Sambruna et al. 2001), rather than the other
models invoking the inverse Compton scattering of electrons.
The size of the X-ray halo, where a large amount of energetic particles is
probably drifting, appears to larger than that of the radio hot spot
(R00; M01), implying that the spatioscale of confinement region of
accelerated particles tends to be larger than that of shock accelerator.

In the knot distant more than $\sim 20~{\rm kpc}$ from core (B95),
the flow is decelerated to the weak relativistic speed of
$(0.21\pm 0.04)c$ (M89), where $c$ is the speed of light, although
in the vicinity of core, superluminal motion of ejecta is observed.
The flow may be collimated by magnetic fields (HH02), resulting in no
significant radial expansion.
Polarization measurements suggest the ordered magnetic field, and
its strength is of the order of sub-milligauss at this knot (M89; R00).
The degree of polarization increases toward the direction of
the bending extended structure, called ``inner extension'' (RM91),
indicating that the field lines are inclined with respect to the jet axis.

\subsection{\it Particle Acceleration by an Oblique Shock}

For knot~A1, presuming the shock speed to be nonrelativistic is adequate
for modeling the DSA based on the Fermi-I mechanism (Gaisser 1990: G90).
We consider the fast-mode oblique shock, for which magnetic field strength
is boosted to $B_{2}=\sqrt{\cos^{2}\theta_{1}+r^{2}\sin^{2}\theta_{1}}B_{1}$,
where the subscripts $i=$'1' and '2' indicate the upstream and downstream
region, $\theta_{i}$ denotes the inclination angle of mean magnetic field line
with respect to the direction normal to shock surface, and $r$ is the
shock compression ratio.
In contrast with the case of the parallel shock with
$\theta_{1}=\theta_{2}=0^{\circ}$ in which particles are deflected due
solely to scattering by magnetic field fluctuations, in the
$\theta_{1}\neq 0^{\circ}$ case a fraction of the particles
is directly reflected at the shock front by the boosted field in the
region '2', conforming to the conservation of magnetic moment.
Expected here is that effect of this mirror-reflection leads to significant
reduction of acceleration time, {\it viz}., increase of acceleration
efficiency (e.g., Kirk \& Heavens 1989: KH89).
Indeed, this effect has been confirmed by the Monte Carlo simulations
for a nonrelativistic shock
(Naito \& Takahara 1995; Ellison, Baring \& Jones 1995).

For calculation of energy gain of particles, we transfer from plasma rest
frames of the regions '1' and '2' to a proper frame where electric field
vanishes (de Hoffmann \& Teller 1950: HT50); then, transform physical
variables back to those in the original frame (O88).
The mean acceleration time can be defined as the cycle time for one
back-and-forth between the region '1' and '2' divided by the energy gain
per encounter with the shock (G90).
An improved calculation including the mirror-reflection effects
yields the following resultant of the mean acceleration time
(Kobayakawa, Honda \& Samura 2002: KHS02):
\begin{equation}
t_{a,{\rm acc}}=\frac{3r_{{\rm g},a}\beta_{a}c}{U_{1}^{2}}
\frac{r\eta_{a}}{r-1}\left\{\cos^{2}\theta_{1}+\frac{\sin^{2}\theta_{1}}
{1+\eta_{a}^{2}}+\frac{r\cos^{2}\theta_{1}+\left[r^{3}\sin^{2}\theta_{1}/
(1+\eta_{a}^{2})\right]}{(\cos^{2}\theta_{1}+r^{2}\sin^{2}\theta_{1})^{3/2}}
\right\},
\end{equation}
for arbitrary species of particle '$a$'.
Here, $U_{1}$ is the shock speed, $\beta_{a}=v_{a}/c\simeq 1$, and $v_{a}$
the speed of the particle, and furthermore, $\eta_{a}=\ell_{a,\|}/r_{g,a}$,
where $\ell_{a,\|}$ the mean free-path (m.f.p.) along the magnetic field line, 
$r_{{\rm g},a}=\beta_{a}\gamma_{a}m_{a}c^{2}/(|q_{a}|B_{1})$ the gyroradius
in the region '1', $\gamma_{a}=(1-\beta_{a}^{2})^{-1/2}$, and $m_{a}$
and $q_{a}$ the rest mass and charge of the particle, respectively.
Note the allowable range of the field inclination angle of
$\theta_{1}\leq\theta_{1,{\rm max}}=\cos^{-1}(U_{1}/c)$, where
$\theta_{1,{\rm max}}$ is called the de Hoffmann-Teller (HT) limit
for oblique shocks (HT50; KH89; KHS02).

In equation~(1), the shock compression ratio is fixed to $r=4$
for a strong nonrelativistic shock.
We compare the magnetic fluctuations involved in the shock to the
Kolmogorov turbulence establishing the spectral intensity of
$I(k)\propto k^{-5/3}$, where $k$ is the wavenumber of the Alfv\'en waves
(e.g., Biermann \& Strittmatter 1987: BS87).
In the circumstances that when the wave-particle resonance
condition, $r_{{\rm g},a}\sim k^{-1}$, is satisfied, the pitch-angle
scattering of particles becomes effective, the m.f.p. can be denoted as
$\ell_{a,\|}\sim [3r_{{\rm g},a}/(2b)](r_{\rm g,max}/r_{{\rm g},a})^{2/3}$.
Here, $r_{\rm g,max}$ defines the maximum resonant gyroradius, and $b$
($\leq 1$) the ratio of turbulent to mean magnetic energy density.
Notice that setting to $b\sim {\cal O}(1)$ makes the local inclination angle
fluctuate around the average $\theta_{i}$.
Nevertheless, this choice is feasible, as long as the trajectory of the
guiding center drift of the particles bounded by the mean field is not
disturbed, that is, the characteristic timescale of the drift in the shock
vicinity is shorter than the coherence time for the resonant scattering.
As shown later, this condition turns out to be fairly satisfied
in the allowable range of $r_{{\rm g},a}\leq r_{\rm g,max}$.
In this aspect, let us set to the critical value, $b=1$, adequate
for simple estimation of the {\it achievable highest} energy of accelerated
particles, and convenient for making a direct comparison
with a previous result in the special case for parallel shock
[BS87, expecting $3b(U_{1}/c)^{2}\sim 1$].

As the jet seems to be {\it nonuniformly filled}, having helical filaments
(LZ01), we allow the hot spot with its radius $R_{\rm HS}$ to contain
{\it multipartite} shock disks, whose radial size each is associated
with the order of maximum turbulent wavelength, $\sim k_{\rm min}^{-1}$.
In order for the resonance to be locked in phase, the gyroradius cannot
exceed $\sim k_{\rm min}^{-1}$, which is limited by $R_{\rm HS}$.
That is, we have the relation of
$r_{{\rm g},a}\leq r_{\rm g,max}\sim k_{\rm min}^{-1}\lesssim R_{\rm HS}$.
In a specific case that the knot feature is identified with a {\it single}
shock, we reproduce $k_{\rm min}^{-1}\sim R_{\rm HS}$, i.e.,
$r_{\rm g,max}\sim R_{\rm HS}$, referred to as the
Hillas criterion (Hillas 1984: H84).
Below we regard, in equation~(1), $r_{\rm g,max}$
as a variable in the range of $\lesssim R_{\rm HS}$.

\subsection{\it The Energy Constraints}

By equating the mean acceleration time (1) with the shortest timescale for
the most severe energy restriction, we can derive the maximum possible energy
of particles, defined as $E_{a,{\rm max}}=\gamma_{a}m_{a}c^{2}$.
For proton ($a$='p'), the time-balance equation can be expressed as
\begin{equation}
t_{\rm p,acc}(\gamma_{\rm p},\theta_{1},r_{\rm g,max})=
\min[t_{\rm p,syn}(\gamma_{\rm p}),t_{\rm p\gamma}(\gamma_{\rm p},u_{\rm rad}),
t_{\rm p,esc}(\gamma_{\rm p},r_{\rm g,max}),t_{\rm sh}].
\end{equation}
Here, $t_{\rm p,syn}\simeq 6\pi m_{\rm p}^{3}c/(\sigma_{{\rm T}}
m_{\rm e}^{2}\gamma_{\rm p}B_{1}^{2})$ defines the cooling time for
synchrotron radiation by the accelerated protons (L92),
where $\sigma_{\rm T}$ is the Thomson cross section.
The loss timescale may be rewritten as
$t_{\rm p,syn}\sim 1\times 10^{14}(10^{11}/\gamma_{\rm p})
(0.7 ~{\rm mG}/B_{1})^{2}~{\rm s}$.
Also, the inelastic collision with photons involving
photopionization can be a competitive energy loss process.
For the present purpose, we employ an approximate expression of
the loss timescale,
$t_{\rm p\gamma}\sim (t_{\rm p,syn}/200)[B_{1}^{2}/(8\pi u_{\rm rad})]$
(for 3C~273; BS87), where $u_{\rm rad}$ stands for average energy density
of target radiation fields.
The above expression scales as
$t_{\rm p\gamma}\sim 1\times 10^{12}(10^{11}/\gamma_{\rm p})
(10^{-8}~{\rm erg~cm}^{-3}/u_{\rm rad})~{\rm s}$.
In addition to the loss due to such elementary processes,
the particle escape itself quenches the acceleration.
The escape time can be estimated as
$t_{\rm p,esc}\simeq 1.5R_{\rm XH}^{2}/
[c\ell_{{\rm p},\|}(\gamma_{\rm p},r_{\rm g,max})]$ (H84),
where $R_{\rm XH}$ ($\gtrsim R_{\rm HS}$) represents the radius of
the X-ray halo confining energetic particles (see Sec.~2.1).
The timescale can be expressed as
$t_{\rm p,esc}\sim 8\times 10^{11}(R_{\rm XH}/\ell_{{\rm p},\|})
(R_{\rm XH}/5~{\rm kpc})~{\rm s}$.
Furthermore, the propagation time of shock through the jet possibly limits the
acceleration, while the radial adiabatic expansion
is {\it less} effective for the self-collimating jet (HH02),
in contrast with spherically expanding supernova remnant shocks (KHS02).
The shock propagation time $t_{\rm sh}$ may be interpreted as the age of
the knot, which is crudely estimated as $\sim L/U_{\rm prop}$, where $L$
represents a distance from core to the knot and $U_{\rm prop}(\sim U_{1})$
the speed of proper motion of the knot; to give the scaling of
$t_{\rm sh}\sim 8\times 10^{12}(L/20~{\rm kpc})(0.25 c/U_{1})~{\rm s}$.

\section{NUMERICAL SOLUTIONS: THE ACHIEVABLE HIGHEST ENERGY OF
ACCELERATED PROTONS}

Now, given $\theta_{1}$, $r_{g,{\rm max}}$, and $u_{\rm rad}$,
we self-consistently solve equation~(2) for $\gamma_{\rm p}$.
Along the explanations mentioned above, at the moment we choose the physical
parameters of 3C~273/A1 as $U_{1}=0.25c$ (M89),
$B_{1}=700~{\rm\mu G}$ (R00), $R_{\rm HS}=2~{\rm kpc}$ (B95),
$R_{\rm XH}=5~{\rm kpc}$ (M01), and $L=20~{\rm kpc}$ (B95).
It is noted that the maximum inclination angle of magnetic field lines reaches
$\theta_{1,{\rm max}}=\cos^{-1}(U_{1}/c)=75.5^{\circ}$, and
$\theta_{2,{\rm max}}=\tan^{-1}(r\tan\theta_{1,{\rm max}})=86.3^{\circ}$
for $r=4$, appropriate for referring to as ``quasiperpendicular shock''. 

In Figure~1 for $\theta_{1}=\theta_{1,{\rm max}}$ (left panel) and $0^{\circ}$
(right panel), we show $E_{\rm p,max}=\gamma_{\rm p}m_{\rm p}c^{2}$
as a function of $r_{\rm g,max}$ ($\sim k_{\rm min}^{-1}$), for some
given values of $u_{\rm rad}$ as a parameter.
In a wide range of smaller $r_{\rm g,max}$ and larger $u_{\rm rad}$,
$E_{\rm p,max}$ is determined by the balance of 
$t_{\rm p,acc}=t_{\rm p\gamma}\propto u_{\rm rad}^{-1}$ in equation~(2)
(dotted curves/lines).
For much smaller $u_{\rm rad}$, $E_{\rm p,max}$ saturates,
being determined by $t_{\rm p,acc}=t_{\rm p,esc}\propto r_{\rm g,max}^{-2/3}$
in a large-$r_{\rm g,max}$ region (soild curve/line),
whereas in a small-$r_{\rm g,max}$ region, by $t_{\rm p,acc}=t_{\rm p,syn}$
for $u_{\rm rad}\lesssim 10^{-10}~{\rm erg~cm^{-3}}$ (dot-dashed curve/line).
Even for the chosen (rather small) value of $L$, $t_{\rm sh}$ cannot
be the shortest timescale in right-hand side of equation~(2).
In the both panels, bottom side of short-dashed line:
$r_{\rm g,p}\leq r_{\rm g,max}$ and left side of long-dashed line:
$r_{\rm g,max}\lesssim R_{\rm HS}=2~{\rm kpc}$ indicate the allowable
and plausible region, respectively.
Note that a point of intersection of these two lines, which represents
the Hillas criterion, $r_{\rm g,p}=r_{\rm g,max}=R_{\rm HS}$,
is in a marginal region above the solid curve/line.
The region of $r_{\rm g,p}\gg r_{\rm g,max}$ far above the
short-dashed line, i.e., $\eta_{\rm p}^{2}\ll 1$ for $b=1$,
can be compared to the ``diffusive limit'' (O88) which violates equation~(1).

In the case of $\theta_{1}=0^{\circ}$, the values of $E_{\rm p,max}$
monotonically decrease as $r_{\rm g,max}$ increases, to exhibit the scaling
of $E_{\rm p,max}\propto r_{\rm g,max}^{-1/2}$ for the cases of
$t_{\rm p,acc}=t_{\rm p\gamma}$ and $t_{\rm p,acc}=t_{\rm p,syn}$, and
$E_{\rm p,max}\propto r_{\rm g,max}^{-2}$ for $t_{\rm p,acc}=t_{\rm p,esc}$.
These tendencies reflect the resonant scattering theory in which longer
coherence time (i.e., longer gyroperiod) leads to lower acceleration
efficiency.
For $\theta_{1}\neq 0^{\circ}$, however, $E_{\rm p,max}$ is found
to be significantly enhanced.
As seen in the case of $\theta_{1}=\theta_{1,{\rm max}}$, the resulting boost
of $E_{\rm p,max}$ is prominent in the region below the short-dashed line.
This property suggests a picture that larger value of
$\eta_{\rm p}\propto (r_{\rm g,max}/r_{\rm g,p})^{2/3}$ in equation~(1)
effectively represents larger anisotropy of diffusion coefficients, i.e.,
for $\eta_{\rm p}^{2}\gg 1$
(corresponding to the ``free crossing limit''; O88),
$\kappa_{\parallel}\gg \kappa_{\perp}$, where
the subscripts refer to the direction of the magnetic field line;
they largely assist the mirror-reflection of particles
(see also Figure~2 below).

We are particularly concerned with the {\it actual} maximum energy of
accelerated proton exceeding $10^{20}~{\rm eV}$.
In the left panel, hatched domain indicates the window where
$E_{\rm p,max}$ exceeds $10^{20}~{\rm eV}$ for the condition of
$r_{\rm g}\leq r_{\rm g,max}\lesssim 2~{\rm kpc}$.
In this domain, it is found that $E_{\rm p,max}$ depends largely on the
value of $u_{\rm rad}$, which involves a large uncertainty responsible
for observation.
At this juncture, we infer the value of $u_{\rm rad}$ from
comparing the observed highest frequency of radiation with the synchrotron
cut-off frequency deduced from the calculated maximum energy of electron.
This is feasible, because the X-ray from knot~A1 is arguably ascribed to
the simple synchrotron emission due to accelerated electrons (Sec.~2.1).
For electron ($a$='e') the method to obtain the maximum possible energy
is so analogous to that explained in Sec.~2.3:
the time-balance equation can be expressed as
$t_{\rm e,acc}(\gamma_{\rm e},\theta_{1},r_{\rm g,max})=\min[t_{\rm e,syn}
(\gamma_{\rm e}),t_{\rm ic}(\gamma_{\rm e},u_{\rm rad})]$.
Here, $t_{\rm e,syn}$ denotes the familiar electron synchrotron timescale,
and $t_{\rm ic}=3m_{\rm e}c/(4\sigma_{\rm KN}\gamma_{\rm e}u_{\rm rad})$
the timescale for the inverse Compton scattering (BS87; L92),
where $\sigma_{\rm KN}$ the Klein-Nishina cross section.
For $\theta_{1}=\theta_{1,{\rm max}}$, we have numerically calculated at
$E_{\rm e,max}=\gamma_{\rm e}m_{\rm e}c^{2}$,
and the synchrotron cut-off frequency by using the expression of
$\nu_{\rm c}\sim (3/16)(e/m_{\rm e}^{3}c^{5})E_{\rm e,max}^{2}B_{1}$ (BS87).
As a consequence, consistency of the calculated $\nu_{\rm c}$
with the observed result $\nu_{\rm c,obs}\gtrsim 10^{17}~{\rm Hz}$ (R00)
is found to require $u_{\rm rad}\lesssim 10^{-8}~{\rm erg~cm^{-3}}$ in the
region of $r_{\rm g,max}\lesssim 2~{\rm kpc}$ (not shown in figure).
Note that the upper limit of $u_{\rm rad}$ is close to
the value inferred from the energy equipartition,
$u_{\rm rad}=B_{1}^{2}/(8\pi)\simeq 2\times 10^{-8}~{\rm erg~cm^{-3}}$.
On the other hand, for $\theta_{1}=0^{\circ}$ we get
$\nu_{\rm c}\gtrsim 10^{14}~{\rm Hz}$ (BS87) in the same region of
$r_{\rm g,max}$, such that $\nu_{\rm c}\ll \nu_{\rm c,obs}$.

For the expected value of $u_{\rm rad}\sim 10^{-8}~{\rm erg~cm^{-3}}$,
we see in Figure~1 that at $r_{\rm g,max}=2~{\rm kpc}$, the actual maximum
energies of proton are $E_{\rm p,max}\sim 2\times 10^{20}~{\rm eV}$ and
$1\times 10^{19}~{\rm eV}$ for $\theta_{1}=\theta_{1,{\rm max}}$ and
$0^{\circ}$, respectively.
For the $\theta_{1}=\theta_{1,{\rm max}}$ case, $E_{\rm p,max}$ always
exceeds the threshold of $10^{20}~{\rm eV}$ in the hatched domain, taking the
value of about $2\times 10^{20}~{\rm eV}$, whereas for $\theta_{1}=0^{\circ}$,
$E_{\rm p,max}$ cannot exceed the threshold, even though taking the peak value
of about $6\times 10^{19}~{\rm eV}$ at $r_{\rm g,max}\approx 90~{\rm pc}$.
It is also remarked that in a possible range of
$u_{\rm rad}\lesssim 10^{-9}~{\rm erg~cm^{-3}}$,
$E_{\rm p,max}\gtrsim 5\times 10^{20}~{\rm eV}$ can be achieved
for $\theta_{1}=\theta_{1,{\rm max}}$.

In Figure~2 for $u_{\rm rad}=10^{-8}~{\rm erg~cm^{-3}}$,
the $\theta_{1}$-dependence of $E_{\rm p,max}$ is shown for
some given $r_{\rm g,max}$ as a parameter.
For $r_{\rm g,max}=0.5~{\rm kpc}$ and $1~{\rm kpc}$, in the whole range of
$\theta_{1}$, $E_{\rm p,max}$ are determined by the balance of
$t_{\rm p,acc}=t_{\rm p\gamma}$, to take their maximum values at the HT limit,
$\theta_{1}=\theta_{1,{\rm max}}$.
For a larger $r_{\rm g,max}=2~{\rm kpc}$ and smaller $0.2~{\rm kpc}$,
this time-balance also governs in a major range of $\theta_{1}$,
except for the regions of $\theta_{1}\lesssim 8^{\circ}$ and
$\gtrsim 49^{\circ}$, where $E_{\rm p,max}$ are determined by
$t_{\rm p,acc}=t_{\rm esc}$ and limited by $r_{\rm g,p}\leq r_{\rm g,max}$
giving a constant maximum, respectively.
Evidently, we find that the values of $E_{\rm p,max}$, which are,
at $\theta_{1}=0^{\circ}$, below the threshold of $10^{20}~{\rm eV}$,
exceed the threshold in large-$\theta_{1}$ region.
For the larger $r_{\rm g,max}$, $E_{\rm p,max}$ appears to be more
enhanced for variation of $\theta_{1}$ from $0^{\circ}$ to
$\theta_{1,{\rm max}}$.

\section{CONCLUSIONS}

Referring to the VLBI observations of helical filaments in the jet of
3C~273, we have estimated the maximum possible energy of
protons diffusively accelerated by an oblique shock at knot~A1.
The hot spot-like feature is regarded as multipartite shock disks, or to
contain a single shock whose radial size can be smaller than the
hot spot radius.
The complementary calculation of maximum electron energy
($E_{\rm e,max}/m_{\rm e}c^{2}\gtrsim 10^{7}$) suggests that
the upper limit of radiation energy density is of
order $u_{\rm rad}\sim 10^{-8}~{\rm erg~cm^{-3}}$.
We conclude that for a possible $u_{\rm rad}$-range,
protons can be accelerated beyond $10^{20}~{\rm eV}$,
safely by the quasiperpendicular shock.
The present method might be applicable for solving the problem of
{\it in situ} acceleration of particles in the other objects, although
this work ignores the effects of particle transport, which may be important
for making a comparison with results of precedent
(Takeda et al. 2003; Abbasi et al. 2004) and future experiments
(at the {\it Pierre Auger Observatory}, the
{\it Extreme Universe Space Observatory}, and so forth).

\clearpage

\begin{figure}
\plotone{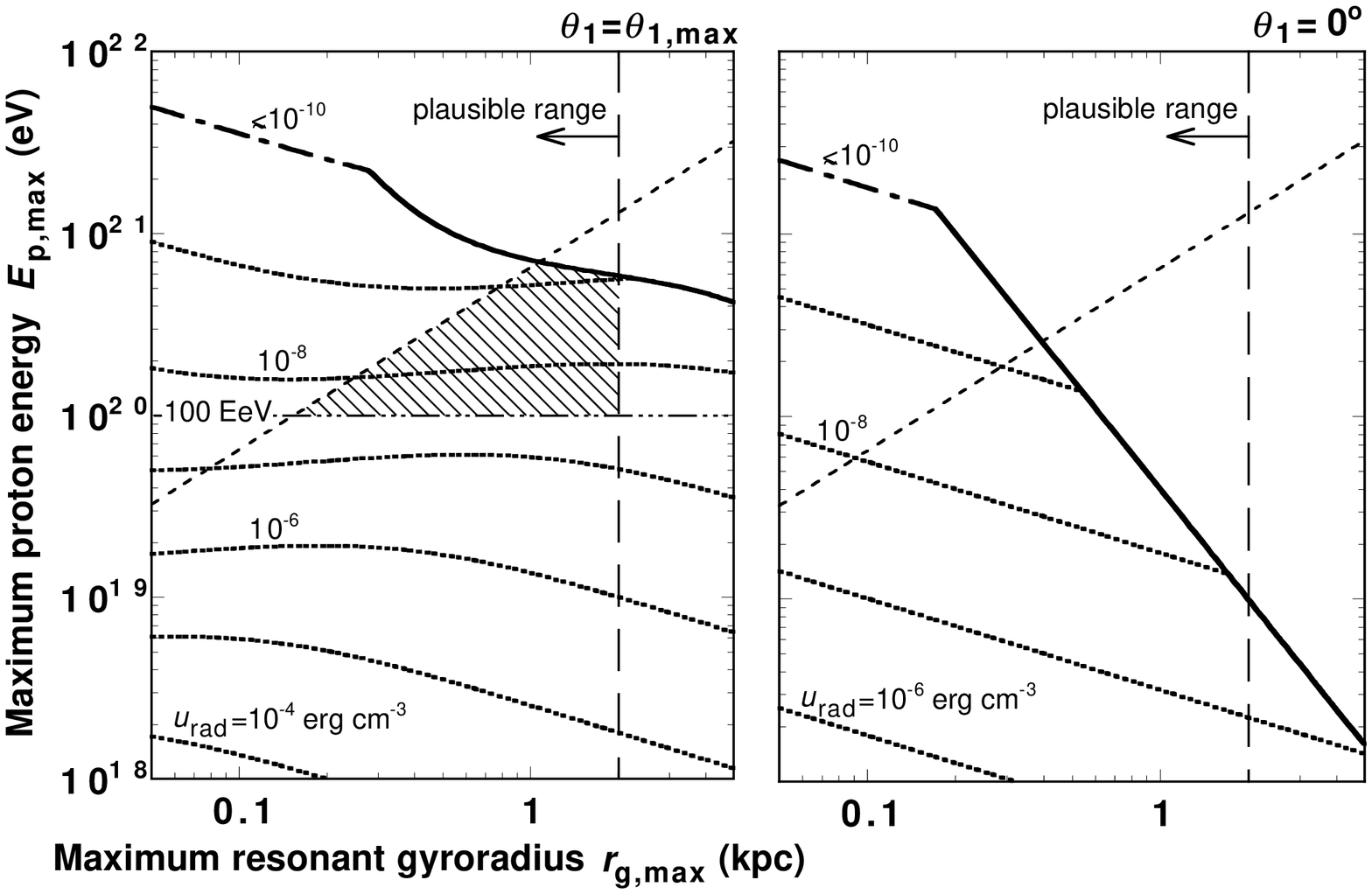}
\vspace{-1cm}
\caption{The maximum possible energy of accelerated proton
$E_{\rm p,max}$ vs. the maximum resonant gyroradius $r_{\rm g,max}$
for the inclination angles of magnetic field lines,
$\theta_{1}=\theta_{1,{\rm max}}=75.5^{\circ}$ (left) and
$\theta_{1}=0^{\circ}$ (right).
The figures have the same axes.
The solutions of equation~(2) are plotted for the balance of the mean
acceleration time, $t_{\rm p,acc}$, with the loss timescales
for escape ($t_{\rm p,esc}$; solid curve/line),
synchrotron ($t_{\rm p,syn}$; dot-dashed curve/line), and
photopionization ($t_{\rm p\gamma}$; dotted curves/lines).
For $t_{\rm p,acc}=t_{\rm p\gamma}$, the curves/lines are displayed each
factor 10 for radiation energy density, $u_{\rm rad}$,
in the unit of ${\rm erg~cm^{-3}}$.
Note that those plots for $u_{\rm rad}\lesssim 10^{-10}~{\rm erg~cm^{-3}}$
overlap the dot-dashed curve/line.
In the left panel, hatched domain indicates the window where
$E_{\rm p,max}$ exceeds the threshold value of $10^{20}~{\rm eV}$
(triple dot-dashed line) for the conditions of $r_{\rm g,p}\leq r_{\rm g,max}$
(bottom side of short-dashed line) and $r_{\rm g,max}\lesssim 2~{\rm kpc}$
(left side of long-dashed line).}
\end{figure}

\begin{figure}
\plotone{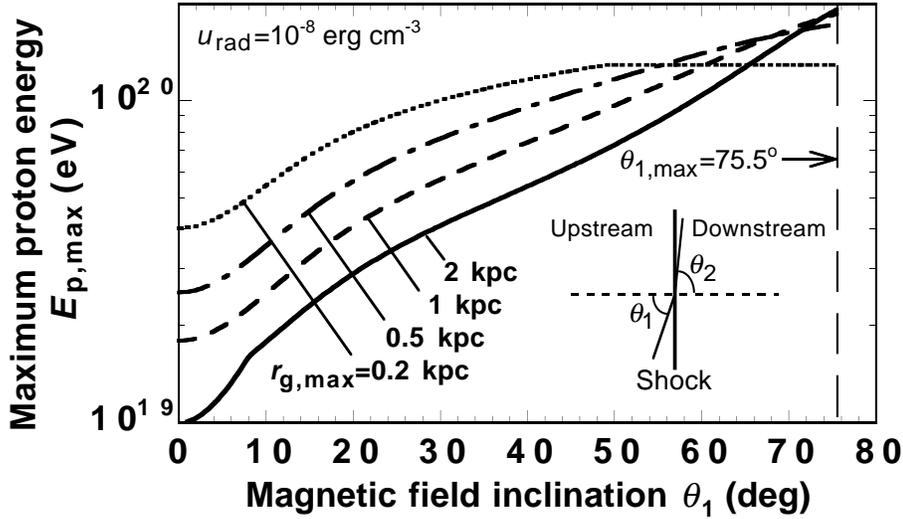}
\vspace{-2.5cm}
\caption{The dependence of the maximum proton energy $E_{\rm p,max}$ on
the inclination angle $\theta_{1}$ for the parameters of $r_{\rm g,max}=0.2$
(dotted), $0.5$ (dot-dashed), $1$ (dashed), and $2~{\rm kpc}$ (solid).
Here, we have set to $u_{\rm rad}=10^{-8}~{\rm erg~cm^{-3}}$.
In the inset, the inclination angles $\theta_{i}$ are limited by the maximum
values of $\theta_{1,{\rm max}}=75.5^{\circ}$ for the shock speed of $0.25c$,
and $\theta_{2,{\rm max}}=86.3^{\circ}$ for the shock compression ratio $4$.}
\end{figure}

\end{document}